%% file: main.tex
\pdfoutput=1
\documentclass{article}

\PassOptionsToPackage{numbers,compress,sort}{natbib}
\usepackage[final]{neurips_2026_vericode}
\workshoptitle{VeriCodeGen: AI for Verifiable Coding}

\usepackage[utf8]{inputenc}
\usepackage[T1]{fontenc}
\usepackage{hyperref}
\hypersetup{hidelinks,
  pdftitle={Forall-Lean-Agent for Auditable Reasoning in Formal Mathematics and Software Verification},
  pdfauthor={Naing Oo Lwin}}
\usepackage{url}
\usepackage{booktabs}
\usepackage{multirow}
\usepackage{amsfonts,amsmath,amssymb}
\usepackage{nicefrac}
\usepackage{microtype}
\usepackage{xcolor}
\input{figure_palette}
\usepackage{enumitem}
\usepackage{graphicx}
\usepackage{caption}
\usepackage{subcaption}
\usepackage{tikz}
\usetikzlibrary{arrows.meta, positioning, calc, fit, matrix, patterns}
\usepackage{pgfplots}
\pgfplotsset{compat=1.18,
  flaChart/.style={
    axis line style={draw=flaMid},
    tick style={draw=flaMid},
    ticklabel style={text=flaNavy},
    label style={text=flaNavy},
    legend style={text=flaNavy},
    every node near coord/.append style={text=flaNavy}
  }
}
\usepackage{array}
\usepackage{makecell}
\usepackage{listings}
\usepackage{tcolorbox}
\usepackage{needspace}
\tcbuselibrary{listings,breakable,skins}
\lstdefinestyle{flaPrompt}{
  basicstyle=\footnotesize\ttfamily,
  columns=fullflexible,
  keepspaces=true,
  breaklines=true,
  breakatwhitespace=true,
  showstringspaces=false,
  aboveskip=0pt,
  belowskip=0pt
}
\newtcblisting{flaPromptBox}[1]{
  enhanced,
  breakable,
  listing only,
  listing options={style=flaPrompt},
  title={#1},
  title after break={#1 (continued)},
  fonttitle=\small\bfseries,
  coltitle=white,
  colbacktitle=flaMid,
  colframe=flaMid,
  colback=flaPale,
  boxrule=0.5pt,
  arc=2mm,
  left=8pt,
  right=8pt,
  top=6pt,
  bottom=6pt,
  toptitle=4pt,
  bottomtitle=4pt,
  before skip=8pt,
  after skip=10pt,
  pad at break*=2mm,
  lines before break=5
}
\newcommand{\fla}{\textsc{Forall-Lean-Agent}}
\newcommand{\vsb}{VeriSoftBench}
\newcommand{\pb}{PutnamBench}
\newcommand{\leval}{Lean Eval}

\title{\fla{} for Auditable Reasoning in\\Formal Mathematics and Software Verification}

\author{%
  Naing Oo Lwin\\
  Astrio Labs\\
  \texttt{nolan.lwin@astriolabs.com}
}
\begin{document}
\maketitle

\begin{abstract}
Coding agents increasingly automate Lean proof development, but successful compilation alone does not establish that a candidate proves the intended statement under acceptable assumptions. We present \fla{}, a frontend-agnostic framework for auditable reasoning in formal mathematics and software verification. The framework combines isolated workspaces, Lean tools, and fresh review with statement comparison, axiom audits, and independent proof checking where supported. Verification evidence and reviewer decisions are bound to the same candidate artifact, making acceptance traceable. We evaluate the framework on \vsb{}, \pb{}, and both problems in the \leval{} software-verification track. On the 100-task \vsb{} subset, integration with \fla{} raises benchmark-rule success from 93 to 100 for GPT-5.6 Sol at low effort while reducing cost from \$69 to \$62. The \pb{} evaluation accepts all 672 problems at an average of \$4.72 each. These results show that agent harness design can improve correctness and efficiency while providing evidence beyond aggregate solve counts.
\end{abstract}

\section{Introduction}

Lean~4 provides a formal language and trusted kernel for checking mathematical arguments and software properties~\citep{demoura2021lean4}. General-purpose coding agents can use this infrastructure to search libraries, construct candidate proofs, and revise them in response to compiler diagnostics. Recent systems demonstrate that this approach can support competition mathematics and substantial formalization tasks without training a dedicated prover~\citep{numina2026,breen2025axprover,requena2026minimal}. Their evaluation depends on the protocol connecting proof generation to verification.

\begin{figure}[t]
\centering
\resizebox{0.89\textwidth}{!}{\input{fig_overview}}
\caption{The \fla{} workflow. A coding agent develops a candidate in an isolated sandbox using Lean and MCP tools. A fresh reviewer recompiles and assesses the candidate without access to the actor's transcript. Acceptance binds the verifier result, reviewer approval, and stored artifact to one digest. Axiom auditing checks dependencies, and \leval{} additionally uses an independent comparator. The input illustrates \vsb{} task 271, \texttt{TM.progress}, with repository context omitted.}
\label{fig:overview}
\end{figure}
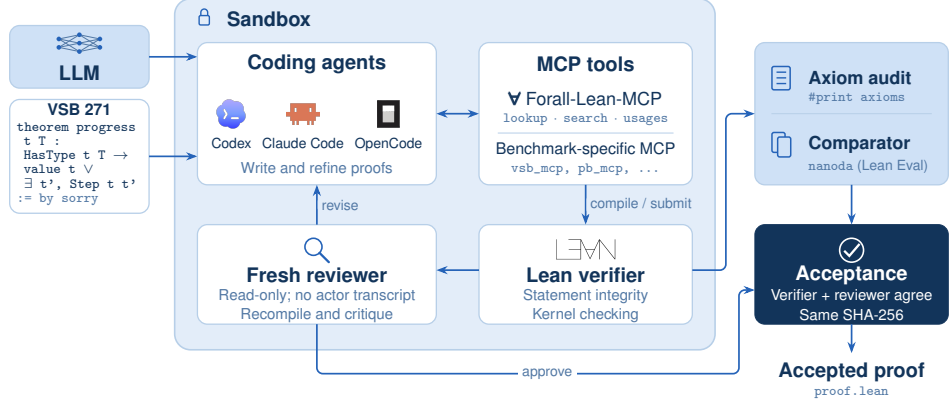

A successful Lean elaboration establishes correctness relative to the elaborated statement and its dependencies. It does not, by itself, establish that an agent preserved the intended statement, avoided reference solutions, or submitted the same artifact that was subsequently reviewed. These distinctions matter for agents with filesystem, shell, and network access. Evaluation also depends on the handling of resource limits and infrastructure errors. A reported failure may reflect an unavailable service or a broken scoring environment, while a reported success may depend on information excluded by the benchmark protocol.

\fla{} addresses these concerns through a common execution and acceptance protocol for Lean proving. It separates an actor that constructs proofs from a fresh reviewer that examines each verified candidate. It restricts runtime access to permitted task context, exposes declaration statements through retrieval tools, and records the evidence used for acceptance. The framework is frontend agnostic, with an adapter interface that supports integration with additional coding agents.

The principal contributions are the following.
\begin{enumerate}[leftmargin=1.4em,itemsep=0.1em,topsep=0.2em]
  \item An auditable execution protocol that combines isolated workspaces, persistent proof generation, fresh review, and acceptance tied to a specific candidate artifact.
  \item An empirical comparison of coding-agent frontends and execution protocols on \vsb-Aristotle, with separate reporting of benchmark success, axiom compliance, and independent-checker coverage.
  \item Applications to mathematical verification on \pb{} and two independently checked \leval{} software developments, together with an analysis of information leakage and infrastructure errors that can invalidate aggregate measurements.

\end{enumerate}

\section{Related work}
\label{sec:related}

\textbf{Agentic proof generation.} COPRA uses execution feedback to guide formal proof generation~\citep{thakur2024copra}. LeanDojo and Pantograph provide interfaces for premise retrieval and interaction with Lean~\citep{yang2023leandojo,pantograph}. More recent approaches place these capabilities within general-purpose agent workflows. Numina-Lean-Agent combines Claude Code with a Lean MCP server for retrieval, compilation, and auxiliary reasoning~\citep{numina2026,mcp2024}. Ax-Prover separates orchestration, proving, and verification~\citep{breen2025axprover}, while \citet{requena2026minimal} study iterative refinement, library search, and context management in a minimal agent. Specialized systems additionally use learned proof search, recursive decomposition, and blueprint construction~\citep{chen2025seedprover,lin2025godel,varambally2025hilbert,goedelarchitect2026,aristotle2025}. \fla{} builds on iterative proof generation and review, with an emphasis on access restrictions and the evidence required to accept a candidate.

\textbf{Evaluation settings.} \pb{} contains 672 formalized competition problems in its Lean answer-given configuration~\citep{tsoukalas2024putnambench}. \vsb{} extends evaluation to 500 proof obligations from 23 repositories, where project-specific definitions and dependencies differ from the Mathlib setting~\citep{verisoftbench2026}. Context dependence is also central to miniCTX~\citep{minictx2025}, while vericoding benchmarks evaluate verified program synthesis~\citep{vericoding2025}. \leval{} includes larger formal developments and checks submissions against trusted challenge statements using a comparator and an independent kernel~\citep{leaneval2026,comparator2026,nanoda}. Recent studies examine code agents for software verification~\citep{codeagentsverif2026,agenticproving2026,goedelcodeprover2026}. Our evaluation uses a shared acceptance design across these settings while retaining their distinct context and budget requirements.

\textbf{Verification and execution control.} Independent comparison of elaborated environments addresses failure modes that compilation alone cannot exclude~\citep{comparator2026}. \fla{} combines verifier-controlled continuation with fresh review, candidate digests, and runtime access logs. Context separation between actor and reviewer is an additional diagnostic mechanism. Formal correctness continues to depend on the verifier and its trusted environment.

\section{Framework}
\label{sec:framework}
\label{sec:method}

\subsection{Execution model}
\fla{} coordinates proof generation, tool access, and acceptance through a task-specific workspace. The actor is a resumable coding-agent session, and the verifier is controlled by the framework. The framework is frontend agnostic. Its adapter interface separates frontend-specific session management from the shared execution and acceptance protocol. Current adapters support Claude Code, Codex, and OpenCode~\citep{claudecode,codexcli,opencode}, and additional frontends can be integrated through the same interface. Benchmark adapters provide the target statement, build environment, and scoring procedure. The separation allows the same review and artifact checks to operate on Mathlib problems, repository theorems, and multi-file developments. The actor cannot modify framework state or acceptance records.

\subsection{Workspace isolation}
\label{sec:blind}
The workspace contains the context permitted by the evaluation protocol. For repository tasks, the actor receives a sanitized project with proof bodies removed and the target theorem withheld from retrieval. For \pb{}, the environment contains the build configuration and compiled Mathlib dependencies without the benchmark checkout that carries reference solutions. We use \emph{runtime blindness} to denote these access restrictions. This term makes no claim about the model's training data.

Shell-enabled configurations enforce filesystem restrictions across the process tree using an operating-system sandbox. A local CONNECT proxy limits outbound traffic to the model provider. Tool-restricted configurations instead validate tool names and resolve path arguments before execution. Restrictions cover alternative checkouts, repository history, previous outputs, and frontend scratch state. Access decisions are logged.

\subsection{Lean tools and retrieval}
\label{sec:tools}
The MCP interface separates general Lean retrieval from benchmark-specific operations. \textbf{Forall-Lean-MCP} provides \texttt{lookup}, \texttt{search}, \texttt{usages}, and \texttt{outline}. Responses contain declaration kinds, statements, and source locations without proof bodies. Benchmark-specific servers, including \texttt{vsb\_mcp} and \texttt{pb\_mcp}, provide task initialization, compilation, and submission. Shell-based adapters can invoke the build and external scorer directly while retaining the acceptance protocol.

Statement-only retrieval makes repository interfaces available without exposing completed proofs through the retrieval channel. Its design was motivated by two exploratory evaluations in which navigation accounted for 51 to 65\% of tool calls and compilation for 10 to 20\%. These exploratory counts motivate the interface without isolating its effect on success.

\subsection{Candidate review and acceptance}
\label{sec:review}
When a candidate passes the verifier, the framework starts a fresh reviewer session. The reviewer receives the task statement and candidate through read-only access and cannot inspect the actor's transcript, previous reviews, or orchestration state. It recompiles the candidate and checks for statement weakening, name shadowing, altered answer values, unsupported assumptions, and vacuous or circular constructions. It then returns an explicit verdict and the candidate digest.

Acceptance requires a verifier pass and reviewer approval for the same artifact. The digest in the submission result must match the reviewer's compilation evidence and the proof stored on disk. A mismatch invalidates acceptance. Rejection feedback is made available to the persistent actor for another revision. An HMAC authenticates resumption state and binds it to the task and directory. Invalid partial state is quarantined.

\subsection{Continuation and termination}
\label{sec:stop}
The framework determines task completion from verification evidence. If an actor turn ends without a verifier pass, the same session resumes with the latest diagnostics and an instruction to continue. External limits on rounds, time, spending, and continuation requests still apply.

The outcome record distinguishes acceptance, exhausted review budgets, resource caps, and infrastructure interruptions. Authentication or usage errors that occur before any inference expenditure leave the task resumable without assigning a proof outcome. Suspected false statements are examined separately from proof failures.

\subsection{Verification and artifact audit}
\label{sec:audit}
\label{sec:verify}
The layered verifier checks candidate syntax, preservation of the frozen statement, declaration shadowing, elaboration, and kernel axioms. The syntactic screen rejects admissions, declared axioms, disallowed imports, and mechanisms such as \texttt{native\_decide}, \texttt{@[implemented\_by]}, and \texttt{@[extern]}. Statement comparison ignores whitespace and comments. The shadowing guard rejects new bindings that change how names in the target resolve, except declarations already present in the trusted template. This matters because redefining \texttt{IsGreatest} as set membership removes its upper-bound requirement while allowing an apparently unchanged theorem to elaborate.

The axiom audit queries \texttt{\#print axioms} for the target. The strict policy permits \texttt{propext}, \texttt{Classical.choice}, and \texttt{Quot.sound}. For \vsb{}, benchmark-rule success is reported separately from strict success because repository lemmas can inherit \texttt{sorryAx} or native-evaluation axioms. A candidate may satisfy the benchmark's scoring rule while failing this stricter policy. \leval{} additionally requires the pinned comparator, \texttt{lean4export}, and the independent \texttt{nanoda} kernel in a Linux environment. Independent-checker coverage on \vsb{} is reported with its own denominator.

Accepted artifacts are re-verified serially against clean repositories. The audit retains the submitted proof, verifier evidence, axiom dependencies, actor and reviewer transcripts, execution metadata, and access logs. Text similarity to withheld reference proofs identifies candidates for manual inspection. Similarity is a screening measure and does not by itself establish leakage or independence. For legacy evaluations in which the final file differs from the submitted file, scoring uses the submit-time artifact recovered from the transcript. Unrecoverable artifacts are excluded.

\section{Experimental setup}
\label{sec:setup}

\textbf{Tasks and coverage.} The main repository evaluation uses the 100-task \vsb-Aristotle subset drawn from 11 repositories. The tasks cover compiler verification, type systems, circuit libraries, probabilistic programs, and separation logic. Task 222 requires a recovered dependency revision because its original pinned revision is no longer available upstream. The replacement targets the same Lean version and is identified separately in the provenance record. The \pb{} evaluation covers 672 problems in the Lean answer-given configuration. The \leval{} evaluation covers both problems in its software-verification track, strong normalization for the calculus of constructions and quantifier elimination for real closed fields.

\textbf{Configurations.} Table~\ref{tab:vsb} lists models, effort, and the target protocol Pass@8 with $r=3$, meaning up to eight samples and three local compiler invocations per sample. Reviewer rounds are separate. Appendices~\ref{app:vsb} and~\ref{app:prompts} document recorded budgets and prompts.

\textbf{Comparisons and accounting.} Within-frontend comparisons fix model and effort, but sampling, compilation, and review budgets differ. We compare benchmark-rule success, strict correctness, tokens, and cost without isolating component effects or assuming equal budgets. Costs include generation, review, and retries. Cumulative scores include documented recovery and subsequent acceptances.

\section{Results}

\subsection{Software verification}
\subsubsection{VeriSoftBench}
\label{sec:vsb}

\begin{table}[t]
\caption{Results on the 100-task \vsb-Aristotle subset. Rules denotes benchmark-rule success, Strict excludes disallowed kernel axioms, and Indep. reports independent-checker acceptances over judged candidates. $^{\S}$Numina-Lean-Agent runs its released protocol, one sample with up to ten rounds, rather than the target protocol used for the other systems, so its total is not a like-for-like comparison. Target protocols are requested settings. Recorded budgets differ. Token counts are in millions. Input totals include cached tokens, shown in parentheses. Output includes reasoning tokens where reported. Dashes indicate unreported values. $^\ddagger$Published baselines~\citep{verisoftbench2026,vsbleaderboard}. Appendix~\ref{app:vsb} documents validation coverage.}
\label{tab:vsb}
\centering
\footnotesize
\setlength{\tabcolsep}{2.5pt}
\begin{tabular}{@{}l l l l r r r r r r@{}}
\toprule
System & Model & Effort & Target protocol & Rules & Strict & Indep. & \makecell{Input tokens\\(cached, M)} & \makecell{Output\\tokens (M)} & Cost \\
\midrule
Aleph Prover$^\ddagger$ & -- & -- & Unknown & 94 & -- & -- & -- & -- & -- \\
Aristotle$^\ddagger$ & -- & -- & Unknown & 69 & -- & -- & -- & -- & -- \\
VeriSoftBench$^\ddagger$ & Gemini-3-Pro & -- & Pass@8, $r{=}3$ & 65 & -- & -- & -- & -- & -- \\
\midrule
Claude Code & Opus 5 & xhigh & Pass@8, $r{=}3$ & 100 & 89 & -- & 53.8 (50.1) & 2.09 & \$135 \\
\fla{} & Opus 5 & xhigh & Pass@8, $r{=}3$ & 100 & 96 & \textbf{69/69} & 45.3 (44.0) & 1.58 & \$111 \\
Claude Code & Opus 5 & low & Pass@8, $r{=}3$ & 99 & 87 & 62/69 & 62.9 (58.7) & 1.18 & \$101 \\
\fla{} & Opus 5 & low & Pass@8, $r{=}3$ & \textbf{100} & 89 & -- & 132.9 (129.4) & 0.95 & \$118 \\

Claude Code & Fable 5 & low & Pass@8, $r{=}3$ & 96 & 89 & 60/66 & 58.0 (53.8) & 1.11 & \$157 \\
\fla{} & Fable 5 & low & Pass@8, $r{=}3$ & \textbf{100} & 89 & 62/70 & 178.3 (172.6) & 1.27 & \$350 \\
\midrule
Numina-Lean-Agent$^{\S}$ & Fable 5 & low & 1 sample, $r{\leq}10$ & 86 & -- & -- & 899.3 (867.8) & 5.15 & \$2,011 \\
\midrule
Codex & GPT-5.6 Sol & low & Pass@8, $r{=}3$ & 93 & 82 & -- & 68.1 (61.7) & 0.95 & \$69 \\
\fla{} & GPT-5.6 Sol & low & Pass@8, $r{=}3$ & \textbf{100} & 88 & 63/69 & 80.7 (75.4) & 0.54 & \$62 \\
Codex & GPT-5.6 Sol & xhigh & Pass@8, $r{=}3$ & 97 & 86 & -- & 63.4 (56.6) & 0.95 & \$68 \\
\fla{} & GPT-5.6 Sol & xhigh & Pass@8, $r{=}3$ & \textbf{100} & 88 & 60/67 & 104.5 (98.1) & 1.00 & \$85 \\
\midrule
OpenCode & GPT-5.6 Sol & low & Pass@8, $r{=}3$ & 67 & 61 & 42/45 & 41.9 (37.1) & 0.54 & \$45 \\
\fla{} & GPT-5.6 Sol & low & Pass@8, $r{=}3$ & 70 & 65 & -- & 26.3 (22.3) & 0.25 & \$33 \\
\bottomrule
\end{tabular}
\end{table}

\begin{figure}[t]
\centering
\begin{subfigure}[t]{0.46\textwidth}
\centering
\begin{tikzpicture}
\begin{axis}[flaChart, ybar, area legend, bar width=9pt, width=\linewidth, height=4.4cm, ymin=88, ymax=102,
  ylabel={\scriptsize tasks solved / 100}, ylabel near ticks,
  symbolic x coords={A,B,C}, xtick=data, enlarge x limits=0.28,
  xticklabels={{Claude Code\\Fable 5, low},{Codex\\Sol, low},{Codex\\Sol, xhigh}},
  xticklabel style={align=center,font=\scriptsize}, yticklabel style={font=\scriptsize},
  nodes near coords, every node near coord/.append style={font=\tiny},
  legend image code/.code={\draw[#1] (0cm,-0.075cm) rectangle (0.15cm,0.075cm);},
  legend style={at={(0.5,1.02)},anchor=south,legend columns=2,font=\scriptsize,draw=none}]
\addplot[fill=flaLine,draw=flaMid] coordinates {(A,96) (B,93) (C,97)};
\addplot[fill=flaBlue,draw=flaNavy] coordinates {(A,100) (B,100) (C,100)};
\legend{{unaided},{with \fla{}}}
\end{axis}
\end{tikzpicture}
\caption{Matched frontend, model, and effort. Blue bars denote \fla{} paired with the named frontend.}
\label{fig:vsb-pairs}
\end{subfigure}\hfill
\begin{subfigure}[t]{0.52\textwidth}
\centering
\begin{tikzpicture}
\begin{axis}[flaChart, xbar stacked, width=0.8\linewidth, height=4.4cm, xmin=0, xmax=100,
  bar width=7pt, enlarge y limits=0.18,
  xlabel={\scriptsize \% of tasks}, xlabel near ticks,
  symbolic y coords={OCt,CX,CL,FL,OP}, ytick={OCt,CX,CL,FL,OP},
  yticklabels={{OpenCode + \textsc{Forall}\\Sol, low},{Codex + \textsc{Forall}\\Sol, xhigh},{Codex + \textsc{Forall}\\Sol, low},{Claude Code + \textsc{Forall}\\Fable 5, low},{Claude Code + \textsc{Forall}\\Opus 5, low}},
  yticklabel style={font=\scriptsize,align=right}, xticklabel style={font=\scriptsize},
  legend style={at={(0.5,1.02)},anchor=south,legend columns=6,font=\tiny,draw=none, /tikz/every even column/.append style={column sep=3pt}}]
\addplot[fill=flaNavy,draw=none] coordinates {(30,FL) (91,CL) (73,CX) (59,OCt) (71,OP)};
\addplot[fill=flaBlue,draw=none] coordinates {(21,FL) (5,CL) (22,CX) (6,OCt) (20,OP)};
\addplot[fill=flaMid,draw=none] coordinates {(11,FL) (3,CL) (1,CX) (5,OCt) (5,OP)};
\addplot[fill=flaLine,draw=none] coordinates {(15,FL) (0,CL) (0,CX) (0,OCt) (4,OP)};
\addplot[fill=flaIce,draw=none] coordinates {(6,FL) (0,CL) (0,CX) (0,OCt) (0,OP)};
\addplot[fill=flaPale,draw=none,postaction={pattern=north east lines,pattern color=flaMid}] coordinates {(17,FL) (1,CL) (3,CX) (30,OCt) (0,OP)};
\legend{r1,r2,r3,r4,r5,exh.}
\end{axis}
\end{tikzpicture}
\caption{Initial recorded outcomes before recovery and retries. Exh. denotes exhaustion.}
\label{fig:rounds}
\end{subfigure}
\caption{Repository proof success under alternative execution protocols and the distribution of review rounds.}
\end{figure}
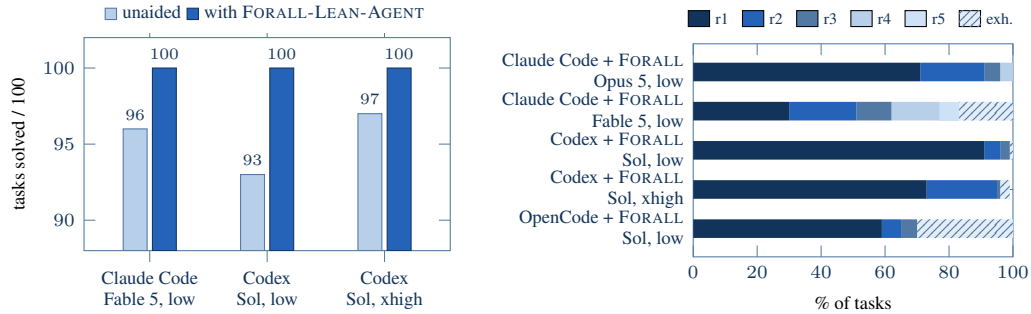

Table~\ref{tab:vsb} reports benchmark success, axiom compliance, independent-checker outcomes, and cost. The results show that model capability and execution design both affect measured performance. The comparisons below separate complete subset evaluations from partial studies and cumulative retry results.

\textbf{Performance without task-specific tools.} Unaided agents solve 93 to 100 of the 100 subset tasks under benchmark rules. Claude Code with Opus~5 solves 100 tasks at \texttt{xhigh} effort and 99 at \texttt{low} effort. Codex solves 93 and 97 at the corresponding effort levels. The published Aleph Prover, Aristotle, and custom in-house Gemini-3-Pro baselines report 94, 69, and 65 tasks, respectively~\citep{verisoftbench2026,vsbleaderboard}. Detailed access conditions for Aleph are unreported. The model, context, and budget differences preclude attributing this gap solely to the framework. At high effort, the tool-assisted and unaided Opus~5 evaluations both approach the benchmark ceiling.

\textbf{Effect of the agent harness.} At fixed low effort, \fla{} + Claude Code improves Fable~5 from 96 to 100 solved tasks, while \fla{} + Codex improves GPT-5.6 Sol from 93 to 100. The \fla{} + Claude Code configuration with Fable~5 solves all four tasks missed by its unaided counterpart. With Opus~5 at low effort, \fla{} improves the unaided configuration from 99 to 100 solved tasks and from 87 to 89 under the strict axiom policy, recovering task 121 after eight unaided samples fail. This configuration reaches the same counts as unaided Opus~5 at \texttt{xhigh}, 100 and 89, at a lower cost of \$118 against \$135, while the larger strict gain observed at \texttt{xhigh}, 89 to 96, does not appear at low effort. \fla{} + Codex at \texttt{xhigh} reaches 100 tasks across the initial evaluation and subsequent retries. Appendix~\ref{app:vsb} records validation coverage.

Several model and harness configurations reach the benchmark-rule ceiling on the 100-task subset, limiting the ability of solve counts to distinguish performance. We did not evaluate the full 500-task \vsb{} benchmark and therefore cannot assess whether this saturation extends to the complete benchmark.

\textbf{Variation across frontends.} With GPT-5.6 Sol at low effort, \fla{} + OpenCode solves 67 tasks without retrieval and 70 with the Lean tool integration, compared with 100 for \fla{} + Codex. The \fla{} + OpenCode no-tool transcripts contain 1,029 reads, 300 searches with \texttt{grep}, and 690 patches. The tool-integrated configuration reduces cost from \$44.84 to \$33.17 and solves three additional tasks. Frontend behavior and tool integration therefore affect performance under the same nominal model and effort setting.

\subsubsection{Lean Eval}
\label{sec:leval}

\begin{table}[!b]
\caption{\leval{} software-verification developments completed with \fla{} and GPT-6 Astra. The submissions pass native compilation, fresh review, comparator validation, and independent-kernel checks.}
\label{tab:leval}
\centering
\footnotesize
\setlength{\tabcolsep}{2.5pt}
\begin{tabular}{@{}l l l l l@{}}
\toprule
Problem & Holes & Submission & Model & Verification \\
\midrule
\texttt{coc\_strong\_normalization} & 6 thm & 233 files / 28.6k lines & GPT-6 Astra & Benchmark validated \\
\texttt{rcf\_quantifier\_elimination} & 1 def + 3 thm & 1 file / 1.5k lines & GPT-6 Astra & Benchmark validated \\
\bottomrule
\end{tabular}
\end{table}

The \leval{} software-verification tasks require connected formal developments with guards against vacuous implementations. The calculus-of-constructions task contains six obligations covering typing, reduction, subject reduction, strong normalization, and consistency. The real-closed-field task requires a quantifier-elimination definition and three associated theorems. Both accepted candidates pass the benchmark's comparator and independent-kernel checks.

For each review, the framework copies the permitted solver files into a pristine benchmark workspace and binds the review to a digest of the complete candidate. The normalization submission contains 233 files and approximately 28.6 thousand lines of Lean. Its development includes operator assistance with checkpoints, infrastructure recovery, and strategic guidance. The final proof covers all six obligations without a universe bound. It is consequently reported as an assisted development, without a standardized single-attempt cost or time claim. The real-closed-field submission contains one file and approximately 1.5 thousand lines. Its accepted implementation has no stated degree or depth restriction.

The cases assess larger-artifact support without providing a controlled model comparison.

\subsection{Mathematical verification}
\label{sec:pb}

\begin{table}[!b]
\caption{Comparison on the Lean answer-given version of \pb{}. Leaderboard totals~\citep{putnamleaderboard} use different benchmark versions, models, budgets, and statement revisions. Reported resource statistics for \fla{} cover all 672 accepted problems.}
\label{tab:pb}
\centering
\footnotesize
\setlength{\tabcolsep}{4pt}
\begin{tabular}{@{}>{\raggedright\arraybackslash}p{4.2cm} >{\raggedright\arraybackslash}p{3.9cm} r >{\raggedright\arraybackslash}p{3.1cm}@{}}
\toprule
System & Reported resources & Solved & Evaluation setting \\
\midrule
Aleph Prover~\citep{aleph2026} & avg \$74, max \$1{,}468 & 672 & closed system \\
NEAR AI, DeepSeek V4$^{a}$ & mean \$0.17 & 672 & shell agent with retries \\
Humanfia$^{b}$ & avg \$44.5 & 672 & GPT-5.6 Sol at xhigh effort \\
Goedel-Architect~\citep{goedelarchitect2026} & 8 blueprints, 4 retries per node, avg \$5.8 & 642 & \\
Seed-Prover 1.5~\citep{chen2025seedprover} & 10 H20-days & 581 & \\
Hilbert~\citep{varambally2025hilbert} & avg Pass@1840 & 462 & evaluated on 660 problems \\
Minimal agent~\citep{requena2026minimal} & avg \$12.60, 50 iterations & 365 & Claude Opus 4.5 \\
Ax-Prover~\citep{breen2025axprover} & Pass@1, $\approx$100 tool calls & 91 & \\
\midrule
\fla{} & avg \$4.72 & \textbf{672} & Opus 5, xhigh \\

\bottomrule
\end{tabular}
\vspace{3pt}
\begin{minipage}{\textwidth}
\scriptsize
$^{a}$\url{https://github.com/SkidanovAlex/putnambench-deepseek}\\
$^{b}$\url{https://github.com/humanfia/putnambench}
\end{minipage}
\end{table}

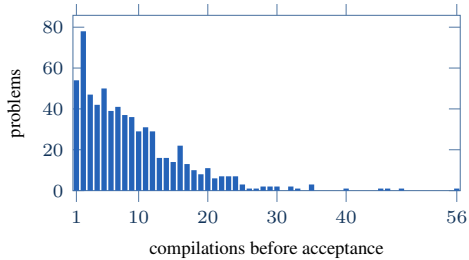
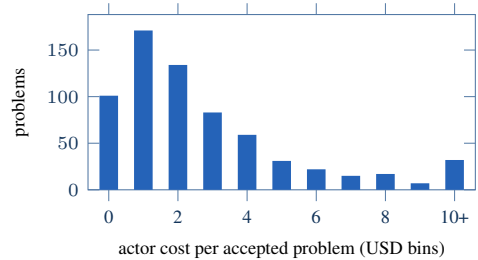
\begin{figure}[!b]
\centering
\begin{subfigure}[t]{0.48\textwidth}
\centering
\begin{tikzpicture}
\begin{axis}[flaChart, ybar, bar width=2pt, width=\linewidth, height=3.9cm, xmin=0.5, xmax=56.5, ymin=0,
  xlabel={\scriptsize compilations before acceptance}, ylabel={\scriptsize problems}, xlabel near ticks, ylabel near ticks,
  xtick={1,10,20,30,40,56}, ticklabel style={font=\scriptsize}]
\addplot[fill=flaBlue,draw=none] coordinates {(1,54)(2,78)(3,47)(4,42)(5,50)(6,39)(7,41)(8,37)(9,36)(10,29)(11,31)(12,29)(13,16)(14,16)(15,14)(16,22)(17,13)(18,10)(19,8)(20,11)(21,6)(22,7)(23,7)(24,7)(25,3)(26,1)(27,1)(28,2)(29,2)(30,2)(32,2)(33,1)(35,3)(40,1)(45,1)(46,1)(48,1)(56,1)};
\end{axis}
\end{tikzpicture}
\caption{Compilations with median 7, mean 8.9, and maximum 56 across 672 accepted problems.}
\end{subfigure}\hfill
\begin{subfigure}[t]{0.48\textwidth}
\centering
\begin{tikzpicture}
\begin{axis}[flaChart, ybar, bar width=7pt, width=\linewidth, height=3.9cm, xmin=-0.6, xmax=10.6, ymin=0,
  xlabel={\scriptsize actor cost per accepted problem (USD bins)}, ylabel={\scriptsize problems}, xlabel near ticks, ylabel near ticks,
  xtick={0,2,4,6,8,10}, xticklabels={0,2,4,6,8,10+}, ticklabel style={font=\scriptsize}]
\addplot[fill=flaBlue,draw=none] coordinates {(0,101)(1,171)(2,134)(3,83)(4,59)(5,31)(6,22)(7,15)(8,17)(9,7)(10,32)};
\end{axis}
\end{tikzpicture}
\caption{Actor cost with median \$2.49, mean \$3.52, and maximum \$34.01.}
\end{subfigure}
\caption{Resource use across all 672 accepted \pb{} problems. 662 receive approval in the first review round and 651 require no continuation request. Median wall time is 27 minutes and the maximum is 211 minutes.}
\label{fig:pb}
\end{figure}

We evaluate \fla{} on \pb{} using Opus~5 at \texttt{xhigh} effort. Table~\ref{tab:pb} reports all 672 problems accepted. The resource analysis covers every accepted problem.

662 of 672 accepted candidates pass review in the first round, with the remaining 10 accepted in a second review round. Compilation remains iterative, with a median of 7 invocations before accepted submission and a maximum of 56. Mean actor and reviewer costs are \$3.52 and \$1.20, an average of \$4.72 per accepted problem and \$3{,}169 in total.

\textbf{Scope of inference.} The reported result is a single run at one model and effort setting, not a pass rate over repeated trials, so it does not estimate variance. Acceptance requires verifier success and reviewer approval on one digest-bound artifact; it does not certify that a proof is idiomatic or minimal. Matched controls are needed to separate the contributions of retrieval, review, and continuation. The evaluation also spans two billing accounts and one harness revision, neither of which changes the model, effort setting, or acceptance rule.

\subsection{Interpreting performance and efficiency}

\textbf{Closed-source comparisons.} Published scores from closed-source systems provide useful reference points, but direct comparison is difficult when model details, tool access, inference budgets, or cost accounting are unavailable. Aggregate scores alone cannot establish whether a difference arises from model capability or the surrounding agent harness. We therefore distinguish published leaderboard results from matched comparisons within our evaluation.

\textbf{Benchmark ceilings and saturation.} At the subset ceiling, strict correctness and resource consumption remain informative. For example, Opus~5 at \texttt{xhigh} solves 100 tasks with and without \fla{}, while strict successes increase from 89 to 96 and total cost falls from \$135 to \$111.

\textbf{Cost across execution protocols.} With Fable~5 at low effort, Numina-Lean-Agent solves 86 tasks for \$2{,}011, compared with 100 for \$350 using \fla{} and 96 for \$157 unaided. Cost per solved task is \$23.38, \$3.50, and \$1.64, respectively. Numina uses one sample with up to ten rounds and uncapped compilation. These protocol differences limit attribution of the observed cost gap to stopping behavior or any individual harness component.

\textbf{Model capability and harness efficiency.} Strong performance without task-specific tools demonstrates the contribution of the underlying model. The harness determines how that capability is used through context selection, tool access, feedback, and review. Its efficiency gains depend on the configuration. With GPT-5.6 Sol at low effort, \fla{} increases benchmark-rule successes from 93 to 100 while reducing cost from \$69 to \$62, corresponding to \$0.62 per solved task. With Fable~5, successes increase from 96 to 100, but cost rises from \$157 to \$350 and strict successes remain at 89. The results support configuration-dependent gains in efficiency, rather than a uniform reduction in inference cost.

\textbf{Axiom audit and independent checking.} The axiom audit inspects the assumptions on which a compiled theorem depends. Strict success requires compliance with the permitted axiom policy described in Section~\ref{sec:audit}. Independent checking additionally validates the candidate against the benchmark statement and replays its exported proof with a separate kernel implementation. The Indep. column reports accepted candidates over those judged by this checker, not over all 100 tasks. These checks assess different properties, and incomplete independent-checker coverage should not be interpreted as either success or failure on the remaining tasks.

\section{Evaluation integrity}
\label{sec:integrity}

\begin{table}[t]
\caption{Observed threats to evaluation validity and the corresponding controls. Access violations can inflate success counts, while artifact drift, scorer failures, and service interruptions can produce incorrect negative outcomes.}
\label{tab:integrity}
\centering
\footnotesize
\setlength{\tabcolsep}{4pt}
\begin{tabular}{@{}p{4.3cm} p{4.0cm} p{5.1cm}@{}}
\toprule
Threat & Evidence & Control \\
\midrule
Permission bypass allowed access to the reference repository & reference similarity of 0.99 to 1.00 and transcript path evidence & operating-system restrictions across the process tree \\
Unrestricted network access exposed public reference material & reference similarity of 1.00 and recorded \texttt{curl} access & provider-only network proxy with refusal logs \\
Filesystem search located alternative benchmark copies & reference similarity of 1.00 and recorded search paths & pattern-based restrictions covering alternative checkouts \\
Stored and submitted proofs differed in 5 of 97 cases & serial verification of recovered submit-time artifacts & candidate digests and single-use sample directories \\
Scorer dependencies were unavailable for five valid samples & independent rescoring of an all-negative task & pinned interpreter and toolchain with environment checks \\
Authentication errors generated 229 false failure records & error returns with zero inference expenditure & interruption status and resumable tasks \\
\bottomrule
\end{tabular}
\end{table}

The incidents in Table~\ref{tab:integrity} identify three distinct obligations for an auditable evaluation. First, information access must be enforced beyond the agent's prompt. A settings-based deny list was ineffective under permission-bypass execution, and restricting one checkout left other local copies and public network sources accessible. Similarity screening and transcript inspection exposed these violations. A valid runtime-blind result therefore requires both effective access controls and records showing how the candidate was produced.

The network control is load-bearing rather than precautionary. Across the 100-task Numina-Lean-Agent evaluation, whose audit records every connection attempt, 26 of 86 audited tasks issue a total of 425 requests to public code-hosting endpoints during solving, including \texttt{github.com}, \texttt{codeload.github.com} and \texttt{raw.githubusercontent.com}. All are refused, and \texttt{api.anthropic.com} is the only host ever permitted. The benchmark repositories are themselves public, so whether any request would have retrieved reference material is not established; the control removes the need to answer that question.

Second, acceptance must identify an immutable artifact. Reused sample directories caused the proof on disk to differ from the proof submitted during the evaluated session. Binding submission, compilation, and review evidence to the same digest makes such discrepancies detectable. Older results require recovery from the submit-time transcript or exclusion when the artifact cannot be reconstructed.

Third, infrastructure failures must remain distinguishable from proof failures. This recurs. During the Numina evaluation, a spend limit and an expired credential each produced batches of tasks that exhausted their round budget without a single model call, recorded as failures whose stated cause was an unproved declaration, which is the staged stub being graded. Sixteen such records were identified by their zero token consumption and re-measured; all five of one batch then succeeded. A harness that aborts on a turn that spent nothing, rather than recording it, prevents these from entering a result table at all. Missing scorer dependencies, expired credentials, and parser defects produced plausible negative records without corresponding unsuccessful reasoning. Separate status categories, environment checks, and serial artifact validation reduce this ambiguity. 

\section{Limitations}
\label{sec:limits}
The evaluation is descriptive, with no repeated configurations, confidence intervals, or component ablations. The contributions of retrieval, review, and persistence remain unresolved. Limited disclosure by closed-source baselines restricts comparisons of capability and efficiency. Results on the 100-task subset do not establish performance on all 500 \vsb{} tasks.

The reviewer shares the actor's model family, so context separation does not establish independent errors. Approval supplements kernel evidence. Runtime restrictions and similarity audits cannot rule out training-data exposure. Temporal holdouts and withheld solutions would strengthen contamination studies. The unreleased harness limits external reproduction.

\paragraph{Broader impacts.} Automated Lean proving can reduce verification effort, but proofs of inadequate specifications can create unwarranted confidence in unsafe systems. Statement comparison, axiom audits, and human specification review help mitigate this risk.

\section{Conclusion}
\fla{} links proof generation and fresh review to verified artifacts through a frontend-agnostic acceptance protocol. The evaluation demonstrates strong performance on the 100-task \vsb{} subset, accepts all 672 \pb{} problems, and supports both independently checked \leval{} software-verification developments. These results demonstrate applicability across formal mathematics and software verification and the value of reporting accepted artifacts alongside verification policies, information boundaries, and resource budgets.

\begin{ack}
We thank Hung Pham, Todd Schmid, Jules Jacobs, and Wojciech R\'{o}\.{z}owski for their advice and guidance. We also thank George Tsoukalas and Jocelyn Qiaochu Chen for supporting our submissions to PutnamBench and VeriSoftBench, respectively, and for helping verify and review the results. We are grateful to the anonymous reviewers at NeurIPS VeriCodeGen for their thoughtful feedback.

This work received no external funding. The author declares no competing interests.
\end{ack}

\bibliographystyle{plainnat}
\bibliography{references}

\clearpage
\appendix
\section{Repository results and independent validation}
\label{app:vsb}

\begin{table}[htbp]
\caption{Repository-level results. Codex Sol low includes the final acceptance on \texttt{clean}. Codex Sol xhigh shows the initial 97-task total before fresh retries, with its cumulative result in Table~\ref{tab:vsb}. VCV denotes VCV-io, FR denotes lean-formal-reasoning-program, Ark denotes ArkLib, juvix denotes juvix-lean, Leroy denotes the compiler course, Expr denotes LeanExprEvaluator, iris denotes iris-lean, and pcf denotes pcf-lean.}
\label{tab:perrepo}
\centering
\scriptsize
\setlength{\tabcolsep}{3pt}
\begin{tabular}{@{}l rrrrrrrrrrr r@{}}
\toprule
System and configuration & clean & VCV & FR & Ark & loom & juvix & veil & Leroy & Expr & iris & pcf & Total \\
Tasks & 47 & 17 & 7 & 5 & 5 & 5 & 5 & 4 & 3 & 1 & 1 & 100 \\
\midrule
\makecell[l]{Claude Code\\Fable 5 low, unaided} & 43 & 17 & 7 & 5 & 5 & 5 & 5 & 4 & 3 & 1 & 1 & 96 \\
\makecell[l]{Claude Code\\Opus 5 low, unaided} & 46 & 17 & 7 & 5 & 5 & 5 & 5 & 4 & 3 & 1 & 1 & 99 \\
\makecell[l]{\fla{} + Claude Code\\Fable 5 low} & 47 & 17 & 7 & 5 & 5 & 5 & 5 & 4 & 3 & 1 & 1 & 100 \\
\makecell[l]{\fla{} + Claude Code\\Opus 5 low} & 47 & 17 & 7 & 5 & 5 & 5 & 5 & 4 & 3 & 1 & 1 & 100 \\
\makecell[l]{\fla{} + Codex\\Sol low} & 47 & 17 & 7 & 5 & 5 & 5 & 5 & 4 & 3 & 1 & 1 & 100 \\
\makecell[l]{\fla{} + Codex\\Sol xhigh} & 46 & 15 & 7 & 5 & 5 & 5 & 5 & 4 & 3 & 1 & 1 & 97 \\
\makecell[l]{\fla{} + OpenCode\\Sol low, no tools} & 28 & 14 & 3 & 4 & 4 & 5 & 5 & 2 & 0 & 1 & 1 & 67 \\
\makecell[l]{Numina-Lean-Agent + Claude Code\\Fable 5 low} & 45 & 10 & 5 & 4 & 3 & 5 & 5 & 4 & 3 & 1 & 1 & 86 \\
\bottomrule
\end{tabular}
\end{table}

\textbf{Execution budgets.} Model identifiers are \texttt{claude-opus-5}, \texttt{claude-fable-5}, \texttt{gpt-5.6-sol}, and \texttt{gpt-6-astra}. A sample starts a fresh actor session. A reviewer round uses a persistent actor and fresh review after verifier success. A continuation resumes the actor, while a fresh retry starts a new session. Unaided Claude Code and Codex use up to eight samples with three local compiler invocations per sample. The Opus-xhigh MCP evaluation uses one sample with uncapped compilation. Reviewed Claude Code configurations allow up to five rounds with uncapped compilation. Reviewed Codex and OpenCode configurations allow up to three rounds and three local actor compilations per task. Numina allows one sample with up to ten rounds and uncapped compilation. Figure~\ref{fig:rounds} shows initial recorded statuses, while Table~\ref{tab:vsb} includes subsequent recovery and retries. These recorded budgets do not establish that every historical timeout or continuation override is recoverable.

\textbf{Kernel dependencies.} The audit re-elaborates each accepted target with \texttt{\#print axioms}. Twelve subset tasks can inherit disallowed axioms from repository lemmas. Tasks 128, 129, 145, 149, 152, and 153 depend on \texttt{Lean.ofReduceBool} or \texttt{Lean.trustCompiler}. Tasks 266, 268, 272, 275, 454, and 457 can inherit \texttt{sorryAx} from admitted dependencies. Reference proofs show the corresponding dependencies. Some generated proofs avoid these lemmas, so the strict count varies by configuration. No additional agent-introduced axiom was identified in the audited candidates. Unavailable audits remain unreported rather than being assigned a clean result.

\textbf{Independent-checker procedure.} The checker evaluates candidates from 72 tasks in repositories pinned to Lean v4.24.0 or v4.24.0-rc1. These repositories are \texttt{clean}, \texttt{veil}, \texttt{LeanExprEvaluator}, and \texttt{VCV-io}. It builds a trusted \texttt{Challenge} containing the target statement and a separate \texttt{Solution} containing the candidate under \texttt{landrun}. Both environments are exported with \texttt{lean4export}. The comparison requires declarations referenced by the statement to agree, inspects exported axiom dependencies, and checks the solution through Lean and \texttt{nanoda}. The implementation uses comparator revision \texttt{71b52ec} with two compatibility patches for Lean versions earlier than v4.27.

Only solved candidates that the checker can judge enter the independent-checker denominator. One or two candidates per configuration are excluded when module-dependent generated names prevent comparison of otherwise matching contexts. Seven Numina candidates from the earlier 49-task evaluation additionally require repository helpers absent from the per-task environment. Among judged cases, no candidate is rejected on a task whose reference proof passes the same checker. The observed rejections concern inherited repository axioms. This finding is conditional on checker coverage and does not certify the unjudged candidates.

\textbf{Supplementary probes and retries.} The Opus-5-low reviewed configuration is measured over all 100 tasks. Every task is accepted, 71 in the first review round and the remainder within four, at a total billed cost of \$118.41. Its axiom audit reports 89 clean targets and 11 carrying forbidden axioms, all of which the corresponding reference proofs also carry, so no axiom is agent-introduced. \fla{} + Codex at xhigh effort initially accepts 96 of the 99 tasks available under their original dependency pins. Recovered task 222 adds one acceptance, and fresh retries of tasks 140, 452, and 462 complete the cumulative set. The \$85 total includes these continuations. Its independent-checker result in Table~\ref{tab:vsb} covers the original 96 accepted tasks, so its denominator differs from the cumulative benchmark score.

\paragraph{Existing assets.} The retained \vsb{} distribution uses the MIT License, and the Lean portion of \pb{} uses Apache License 2.0. These benchmark licenses do not replace the licenses of their dependencies or the terms governing hosted model access. The submission does not redistribute the benchmark repositories or model weights.

\textbf{Reference similarity.} Raw-text sequence similarity is computed between each accepted candidate and its withheld reference proof. Ratios of 0.99 to 1.00 identify the confirmed access violations discussed in Section~\ref{sec:integrity}. Other high-similarity cases include short proofs and conventional induction patterns. Transcript inspection and reproduction by another model help assess these cases, but neither establishes absence of training-data exposure. The audit is therefore used to prioritize investigation rather than to assign an unconditional independence guarantee.

\section{PutnamBench coverage and validation}
\label{app:pb}

\begin{table}[htbp]
\caption{Year-ordered \pb{} coverage. The evaluation records and accepts all 672 problems.}
\label{tab:pbyears}
\centering
\footnotesize
\setlength{\tabcolsep}{3pt}
\begin{tabular}{@{}l rrrrrrrrrrrrrrrr@{}}
\toprule
Year & 62 & 63 & 64 & 65 & 66 & 67 & 68 & 69 & 70 & 71 & 72 & 73 & 74 & 75 & 76 & 77 \\
In benchmark & 11 & 9 & 12 & 12 & 12 & 12 & 11 & 11 & 10 & 10 & 11 & 9 & 10 & 11 & 9 & 10 \\
Accepted & 11 & 9 & 12 & 12 & 12 & 12 & 11 & 11 & 10 & 10 & 11 & 9 & 10 & 11 & 9 & 10 \\
\midrule
Year & 78 & 79 & 80 & 81 & 82 & 83 & 84 & 85 & 86 & 87 & 88 & 89 & 90 & 91 & 92 & 93 \\
In benchmark & 11 & 10 & 10 & 8 & 9 & 9 & 8 & 10 & 12 & 11 & 12 & 9 & 10 & 10 & 11 & 11 \\
Accepted & 11 & 10 & 10 & 8 & 9 & 9 & 8 & 10 & 12 & 11 & 12 & 9 & 10 & 10 & 11 & 11 \\
\midrule
Year & 94 & 95 & 96 & 97 & 98 & 99 & 00 & 01 & 02 & 03 & 04 & 05 & 06 & 07 & 08 & 09 \\
In benchmark & 11 & 10 & 10 & 9 & 10 & 11 & 10 & 9 & 8 & 12 & 10 & 11 & 10 & 11 & 11 & 11 \\
Accepted & 11 & 10 & 10 & 9 & 10 & 11 & 10 & 9 & 8 & 12 & 10 & 11 & 10 & 11 & 11 & 11 \\
\midrule
Year & 10 & 11 & 12 & 13 & 14 & 15 & 16 & 17 & 18 & 19 & 20 & 21 & 22 & 23 & 24 & 25 \\
In benchmark & 12 & 12 & 11 & 10 & 11 & 12 & 11 & 9 & 12 & 11 & 9 & 10 & 12 & 12 & 11 & 12 \\
Accepted & 12 & 12 & 11 & 10 & 11 & 12 & 11 & 9 & 12 & 11 & 9 & 10 & 12 & 12 & 11 & 12 \\
\bottomrule
\end{tabular}
\end{table}

\textbf{Model and verification.} The \pb{} evaluation uses Opus~5 at \texttt{xhigh} effort with the answer-given statement and Mathlib context. Acceptance requires verification and fresh review of the same candidate. The diagnostics below cover all accepted problems in the completed evaluation.

\textbf{Resource distributions.} Across the 672 accepted problems, compilation count has median 7, mean 8.9, and maximum 56. Actor expenditure has median \$2.49, mean \$3.52, maximum \$34.01, and rounded total \$2{,}364. Reviewer expenditure averages \$1.20 and totals approximately \$805, giving an average of \$4.72 per accepted problem and \$3{,}169 overall. Wall time has median 27 minutes, 90th percentile 68 minutes, and maximum 211 minutes. Actor turns have median 31 and maximum 169. Acceptance required a second review round in 10 cases and a continuation request in 21.

\textbf{Verifier validation.} A fifteen-case test set covers permitted proofs and attempts to bypass the verification policy. The cases include admissions, declared axioms, native evaluation, extra imports, altered statements or answer values, and several forms of name shadowing. Positive controls include comments containing \texttt{sorry}, statement reformatting, and harmless helper declarations. The tests exposed an incorrect rejection of template-defined answer abbreviations and a missed notation-based redefinition. After these cases were addressed, a round-trip check of 1,018 staged candidates recorded no false rejection. This is an empirical check over the retained candidates, not a completeness guarantee for the verifier.

\textbf{Toolchain migration.} The evaluation ran on Lean v4.27.0. After the benchmark moved to Lean v4.33.1 on 22 September 2026, all 672 accepted proofs were re-verified against it. 502 compiled unchanged. The other 170 failed only through changed tactic behavior, such as \texttt{simp} normal forms, \texttt{ring\_nf}, and instance synthesis. An agent restricted to proof bodies repaired all 170 for a total of \$89.90. Each repaired file was re-checked for compilation, permitted axioms, and an unchanged statement and answer term. Resource figures above describe the original evaluation and exclude these repairs.

\begingroup
\raggedbottom
\input{prompt_templates}

\clearpage
\endgroup
\end{document}

%% file: figure_palette.tex
\definecolor{flaNavy}{HTML}{12345A}
\definecolor{flaBlue}{HTML}{2563B8}
\definecolor{flaMid}{HTML}{5278A4}
\definecolor{flaLine}{HTML}{B7CFE8}
\definecolor{flaPale}{HTML}{E3EDF9}
\definecolor{flaIce}{HTML}{CFE1F6}

\definecolor{rndA}{HTML}{082740}
\definecolor{rndB}{HTML}{134A73}
\definecolor{rndC}{HTML}{2A6FA3}
\definecolor{rndD}{HTML}{4E8FC0}
\definecolor{rndE}{HTML}{79ADD6}
\definecolor{unaidedC}{HTML}{4E8FC0}
\definecolor{forallC}{HTML}{082740}

%% file: fig_overview.tex
\begin{tikzpicture}[x=1cm,y=1cm,
  font=\sffamily\fontsize{9}{11}\selectfont,text=flaNavy,
  flaCard/.style={draw=flaLine,fill=white,rounded corners=5pt,line width=.65pt},
  flaFlow/.style={-{Stealth[length=2.0mm,width=1.45mm]},draw=flaBlue,line width=.85pt,rounded corners=3pt},
  flaTitle/.style={font=\sffamily\bfseries\fontsize{10}{12}\selectfont},
  flaDetail/.style={font=\sffamily\fontsize{7.5}{9}\selectfont,text=flaMid},
  flaLabel/.style={font=\sffamily\fontsize{7}{9}\selectfont,text=flaMid,fill=flaPale,inner sep=2pt}
]
\path[fill=flaPale,draw=flaLine,rounded corners=8pt,line width=.8pt]
  (2.30,.55) rectangle (11.80,6.72);
\draw[flaBlue,line width=.7pt,rounded corners=.75pt] (2.73,6.26) rectangle (2.91,6.425);
\draw[flaBlue,line width=.7pt] (2.76,6.425)--(2.76,6.48) arc (180:0:.06)--(2.88,6.425);
\node[flaTitle,anchor=west] at (3.10,6.36) {Sandbox};

\path[flaCard,fill=flaIce] (-.60,5.15) rectangle (1.85,6.25);
\begin{scope}[shift={(.625,5.94)},scale=.82]
\foreach \ya in {-.09,.09} {\foreach \yb in {-.18,0,.18} {
  \draw[flaMid,line width=.45pt] (-.44,\ya)--(0,\yb);
  \draw[flaMid,line width=.45pt] (0,\yb)--(.44,\ya);}}
\foreach \ya in {-.09,.09} {\fill[flaBlue] (-.44,\ya) circle (.050);\fill[flaBlue] (.44,\ya) circle (.050);}
\foreach \yb in {-.18,0,.18} {\fill[flaBlue] (0,\yb) circle (.050);}
\end{scope}
\node[flaTitle] at (.625,5.42) {LLM};
\draw[flaFlow] (1.85,5.70)--(2.70,5.70);

\path[flaCard] (-.60,2.95) rectangle (1.85,4.95);
\node[font=\sffamily\bfseries\fontsize{8}{10}\selectfont] at (.625,4.77) {VSB 271};
\node[font=\ttfamily\fontsize{7}{9}\selectfont,anchor=base west,inner sep=0pt] at (-.48,4.40)
  {theorem progress};
\node[font=\ttfamily\fontsize{7}{9}\selectfont,anchor=base west,inner sep=0pt] at (-.48,4.15)
  {\hspace{.12cm}t T :};
\node[font=\ttfamily\fontsize{7}{9}\selectfont,anchor=base west,inner sep=0pt] at (-.48,3.90)
  {\hspace{.12cm}HasType t T $\to$};
\node[font=\ttfamily\fontsize{7}{9}\selectfont,anchor=base west,inner sep=0pt] at (-.48,3.65)
  {\hspace{.12cm}value t $\lor$};
\node[font=\ttfamily\fontsize{7}{9}\selectfont,anchor=base west,inner sep=0pt] at (-.48,3.40)
  {\hspace{.12cm}$\exists$ t', Step t t'};
\node[font=\ttfamily\fontsize{7}{9}\selectfont,anchor=base west,inner sep=0pt,text=flaMid] at (-.48,3.15)
  {:= by sorry};
\draw[flaFlow] (1.85,3.95)--(2.70,3.95);

\path[flaCard] (2.70,3.45) rectangle (6.90,5.95);
\node[flaTitle] at (4.80,5.57) {Coding agents};
\matrix (agentNames) [matrix of nodes,ampersand replacement=\&,
  column sep=.19cm,inner sep=0pt,
  nodes={font=\sffamily\fontsize{7}{9}\selectfont,anchor=base,
    inner sep=0pt,text height=1.6ex,text depth=.4ex}]
  at (4.80,4.19) {Codex \& Claude Code \& OpenCode \\};
\node at ([yshift=.60cm]agentNames-1-1.base) {\includegraphics[trim=74bp 74bp 74bp 74bp,clip,height=.51cm]{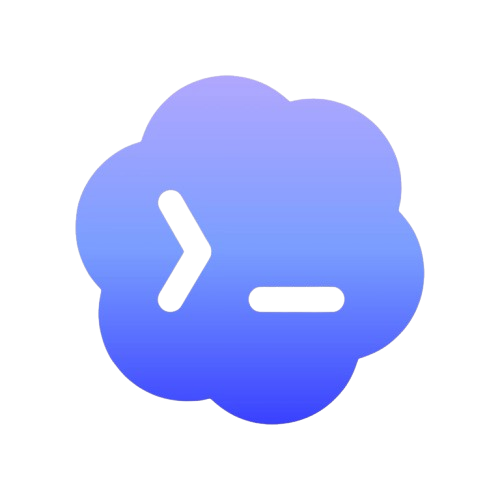}};
\node at ([yshift=.60cm]agentNames-1-2.base) {\includegraphics[trim=0bp 74bp 0bp 92bp,clip,width=.59cm]{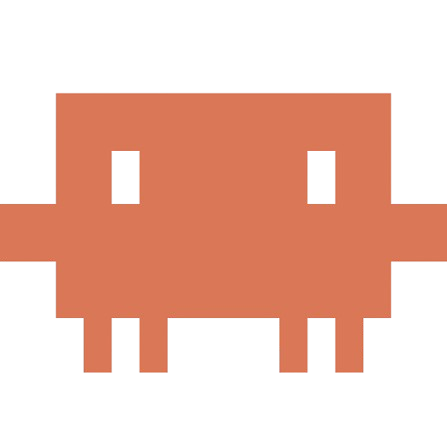}};
\node at ([yshift=.60cm]agentNames-1-3.base) {\includegraphics[trim=68bp 48bp 70bp 48bp,clip,height=.49cm]{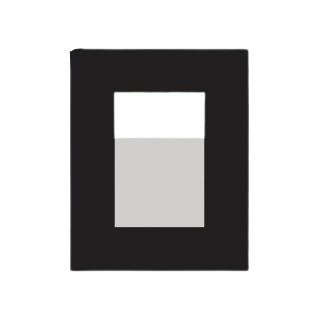}};
\node[flaDetail] at (4.80,3.73) {Write and refine proofs};

\path[flaCard] (7.65,3.45) rectangle (11.40,5.95);
\node[flaTitle] at (9.525,5.57) {MCP tools};
\node[font=\sffamily\fontsize{9}{11}\selectfont] at (9.525,4.97) {$\pmb{\boldsymbol{\forall}}$ Forall-Lean-MCP};
\node[flaDetail,font=\ttfamily\fontsize{7}{9}\selectfont] at (9.525,4.60) {lookup\;\textperiodcentered\;search\;\textperiodcentered\;usages};
\draw[flaLine,line width=.45pt] (7.98,4.36)--(11.07,4.36);
\node[font=\sffamily\fontsize{8}{10}\selectfont] at (9.525,4.07) {Benchmark-specific MCP};
\node[flaDetail,font=\ttfamily\fontsize{7}{9}\selectfont] at (9.525,3.72) {vsb\_mcp, pb\_mcp, \ldots};
\draw[flaFlow,{Stealth[length=2.0mm,width=1.45mm]}-{Stealth[length=2.0mm,width=1.45mm]}] (6.90,4.70)--(7.65,4.70);

\path[flaCard] (7.65,1.00) rectangle (11.40,2.75);
\node at (9.525,2.31) {\includegraphics[trim=221bp 100bp 229bp 21bp,clip,width=1.04cm]{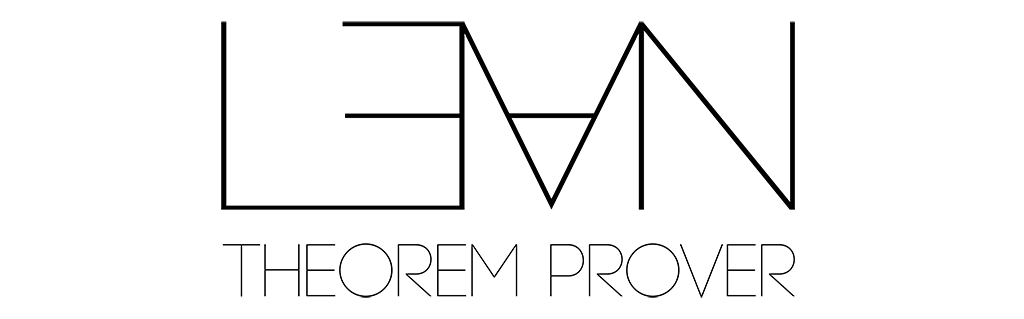}};
\node[flaTitle] at (9.525,1.86) {Lean verifier};
\node[flaDetail] at (9.525,1.49) {Statement integrity};
\node[flaDetail] at (9.525,1.16) {Kernel checking};
\draw[flaFlow] (9.525,3.45)--node[flaLabel,right] {compile / submit} (9.525,2.75);

\path[flaCard] (2.70,1.00) rectangle (6.90,2.75);
\begin{scope}[shift={(4.80,2.30)}]
\draw[flaBlue,line width=.8pt] (-.06,.06) circle (.15);
\draw[flaBlue,line width=1.15pt,line cap=round] (.046,-.046)--(.21,-.21);
\end{scope}
\node[flaTitle] at (4.80,1.86) {Fresh reviewer};
\node[flaDetail] at (4.80,1.49) {Read-only; no actor transcript};
\node[flaDetail] at (4.80,1.16) {Recompile and critique};
\draw[flaFlow] (7.65,1.95)--(6.90,1.95);
\draw[flaFlow] (4.80,2.75)--node[flaLabel,right] {revise} (4.80,3.45);

\path[flaCard,fill=flaIce] (12.50,3.45) rectangle (15.90,5.95);
\draw[flaBlue,line width=.75pt,rounded corners=1pt] (12.80,5.12) rectangle (13.12,5.53);
\foreach \yy in {5.43,5.33,5.23} {\draw[flaBlue,line width=.7pt] (12.88,\yy)--(13.05,\yy);}
\node[font=\sffamily\bfseries\fontsize{9}{11}\selectfont,anchor=west] at (13.32,5.34) {Axiom audit};
\node[flaDetail,anchor=west,font=\sffamily\fontsize{7}{9}\selectfont] at (13.32,4.97) {\texttt{\#print axioms}};
\draw[flaLine,line width=.45pt] (12.79,4.70)--(15.61,4.70);
\draw[flaBlue,line width=.75pt,rounded corners=1pt] (12.80,4.02) rectangle (13.03,4.37);
\draw[flaBlue,line width=.75pt,rounded corners=1pt,fill=flaIce] (12.90,3.94) rectangle (13.13,4.29);
\node[font=\sffamily\bfseries\fontsize{9}{11}\selectfont,anchor=west] at (13.32,4.17) {Comparator};
\node[flaDetail,anchor=west,font=\sffamily\fontsize{7}{9}\selectfont] at (13.32,3.77) {\texttt{nanoda} (Lean Eval)};
\draw[flaFlow] (11.40,1.95)--(11.99,1.95)--(11.99,4.70)--(12.50,4.70);

\path[draw=flaNavy,fill=flaNavy,rounded corners=5pt] (12.50,1.00) rectangle (15.90,2.75);
\draw[white,line width=.8pt] (14.20,2.30) circle (.21);
\draw[white,line width=1.15pt] (14.07,2.29)--(14.16,2.20)--(14.34,2.41);
\node[flaTitle,text=white] at (14.20,1.86) {Acceptance};
\node[flaDetail,text=white] at (14.20,1.49) {Verifier + reviewer agree};
\node[flaDetail,text=white] at (14.20,1.16) {Same SHA-256};
\draw[flaFlow] (14.20,3.45)--(14.20,2.75);
\draw[flaFlow] (4.80,1.00)--(4.80,.13)--(12.22,.13)--(12.22,1.60)--(12.50,1.60);
\node[flaLabel,fill=white] at (8.85,.13) {approve};
\draw[flaFlow] (14.20,1.00)--(14.20,.49);
\node[font=\sffamily\bfseries] at (14.20,.16) {Accepted proof};
\node[flaDetail,font=\ttfamily\fontsize{7}{9}\selectfont] at (14.20,-.19) {proof.lean};
\end{tikzpicture}%

%% file: prompt_templates.tex
\section{Harness prompt templates}
\label{app:prompts}

The following templates reproduce the task instructions supplied by the retained harness implementations. Line wrapping, dash typography, and one incidental remark are edited for presentation. Placeholders denote the task identifier, problem name, candidate digest, or reviewer report inserted at runtime. Task files and tool responses provide the formal statement, repository context, and compiler diagnostics separately. These templates do not include frontend system instructions or task-specific continuation overrides.

Table~\ref{tab:vsb} reports the target protocol, token consumption, and cost of the repository evaluations. Appendix~\ref{app:pb} reports the mathematical evaluation's resource distributions. The prompts below complement this accounting with the instructions used for proof generation, fresh review, and revision.

\Needspace{10\baselineskip}
\subsection{VeriSoftBench}

The actor can initialize the task, inspect and edit its candidate, compile, and submit through the benchmark tools. The reviewer receives read-only inspection and reviewer-specific compilation tools. The task identifier replaces \texttt{<task-id>}.

\begin{flaPromptBox}{Actor}
# Solve VeriSoftBench task <task-id> as the actor

Call `vsb_start` with task id `<task-id>`. It returns the exact theorem statement and
prepares the proof-elided `./tree/`.

If `./REVIEW.md` exists, read it before doing further proof work. It was written
by a fresh, independent, read-only reviewer. Treat its findings as hypotheses:
check them against Lean and address every valid blocking issue.

Read and search `./tree/`, and maintain the candidate proof body in
`./proof.lean`. The proof must begin with `:= by`; never restate or weaken the
theorem. Use `vsb_compile` for deterministic feedback. Call `vsb_submit` only
when the candidate compiles and you believe it is complete.

The outer runner may resume this same actor session after another independent
review. Keep useful reasoning and failed approaches in your session, while the
workspace carries the current candidate between rounds.

Nothing outside this directory and no network access are permitted. Do not try
to alter `REVIEW.md`, transcripts, audit files, metadata, or result files.
\end{flaPromptBox}

\begin{flaPromptBox}{Fresh reviewer}
# Independently review VeriSoftBench task <task-id>

You are a fresh critic. You have no solver conversation and must form your own
judgment from the formal task and current artifact.

1. Call `vsb_review_start` with task id `<task-id>` to obtain the exact theorem
statement
   without changing solver state.
2. Read `./proof.lean` and any relevant definitions or theorem statements in
   the proof-elided `./tree/`.
3. Call `vsb_review_compile` exactly once after inspection. Compiler success is
   required for approval and records evidence bound to the candidate's hash.
4. Check that the proof addresses the exact statement, closes every goal, and
   does not rely on a forbidden escape, hidden admission, statement mutation,
   or irrelevant generated declaration.
5. Do not edit any file and do not call actor compile/start or submission tools.

Return exactly one verdict line followed by concise evidence and, on rejection,
specific next steps. Also return exactly one digest line copied from the
successful `vsb_review_compile` result:

`REVIEW_VERDICT: APPROVE`

or

`REVIEW_VERDICT: REJECT`

`REVIEW_PROOF_SHA256: <64 lowercase hexadecimal characters>`

Be skeptical but evidence-driven. Never approve a candidate that did not pass
`vsb_compile`. Do not reject a kernel-checked proof merely for style.
\end{flaPromptBox}

\begin{flaPromptBox}{Actor continuation}
A fresh independent reviewer has assessed the current candidate. Read ./REVIEW.md,
verify every finding yourself, improve ./proof.lean, compile it, and submit only when
complete.
\end{flaPromptBox}

\Needspace{10\baselineskip}
\subsection{PutnamBench in the answer-given configuration}

The actor receives the frozen theorem and its answer value through \texttt{pb\_start}. The reviewer reads the same task and the candidate through its own tools. The quoted instructions encourage persistence. Their statements about unlimited resources do not replace the external stopping conditions enforced by the runner, as described in Section~\ref{sec:stop}.

\begin{flaPromptBox}{Actor}
Prove the theorem in `<problem-name>.lean`, in this directory.

Use the Forall tools, not the shell, for this task:

  * `pb_start("<problem-name>")`  - begin, and see the task. Call it first.
  * `pb_compile()`        - elaborate your file. UNLIMITED: there is no budget
                            and no penalty. Compile after every single step.
  * `pb_submit()`         - score the file with the five-layer verifier.
  * `forall_lean` search / lookup / usages - retrieval over Mathlib. Prefer it
                            to grepping the source by hand.

`pb_submit` IS THE ONLY THING THAT ENDS THIS TASK. Nothing else counts as
finishing - not a clean `pb_compile`, and not your own judgement that the proof
is done. Submitting is free and unlimited: a failure returns the verifier's
exact words and costs you nothing.

So: WHENEVER YOU BELIEVE THE PROOF IS COMPLETE, CALL `pb_submit()`. If you
submitted earlier while the proof was still unfinished, that submission is dead
- you must call `pb_submit()` again now that it is finished. Never end a turn
having compiled cleanly without submitting.

WHAT COUNTS AS DONE. Not your judgement -- the verifier's, via `pb_submit`.
It checks all of:

  * no `sorry`, `admit`, `native_decide`, `@[implemented_by]`, `@[extern]`, and
    no `axiom` declaration anywhere in the file;
  * no import other than `import Mathlib`;
  * the theorem statement is character-for-character the one you were given
    (whitespace and comments aside), and so is the answer value in the `abbrev` above
    it -- it was given to you,
and changing it is tampering, not solving;
  * no declaration of yours shadows a name the statement resolves through --
    defining your own `IsGreatest` beside a statement that uses Mathlib's is a
    rejection, not a proof;
  * the file elaborates with no errors;
  * `#print axioms` for the theorem shows nothing beyond `propext`,
    `Classical.choice` and `Quot.sound`.

You will see the verifier's exact output when it refuses, so work from what it
says rather than guessing.

HOW LONG YOU HAVE. There is no turn limit and no token budget. Compile as often
as you want -- there is no per-compile cost to you and no penalty for a failed
attempt. Long proofs are expected: budget for hundreds of turns on a hard
problem, and do not try to land it in one shot.

HOW TO WORK. Write a short proof sketch first, in natural language, and only
then formalise it. Turn each step of the sketch into a `have ... := by sorry`
and discharge them one at a time, compiling after each, so you always know
which step is the one that is failing. Search Mathlib for the lemma you need
rather than reproving it. If you come to believe the statement is false, say so
explicitly and try to derive `False` from its hypotheses -- that is a real
finding.
\end{flaPromptBox}

\begin{flaPromptBox}{Fresh reviewer}
You are an independent reviewer. You did not write this proof and you cannot
edit it. You have not seen the author's reasoning - only the statement it must
prove and the file it produced.

Call `pb_review_start("<problem-name>")` to see both, and `pb_review_compile()` to
elaborate the candidate yourself. Do not take the author's word, or a previous
verifier PASS, for anything you can check.

Your job is to find the ways this could be wrong while still compiling:

  * Does it prove the FROZEN statement, or a weaker one? Compare the statement
    in the file against the frozen one character by character.
  * Does any declaration in the file shadow a name the statement resolves
    through? A local `def IsGreatest` beside a statement using Mathlib's
    `IsGreatest` guts the problem while compiling cleanly.
  * Where an answer value was given, is it unchanged?
  * Does anything reach past the kernel - an axiom, `native_decide`, a
    `@[implemented_by]`, an extra import?
  * Are there vacuous or circular steps: a hypothesis that is never used
    because it was silently weakened, a `have` that restates the goal?

Then end your reply with exactly one line:

    VERDICT: APPROVE

if the proof is correct and complete, or

    VERDICT: REJECT

if it is not - and above that line say precisely what is wrong, specifically
enough that the author can fix it. Approving a broken proof is the worst
outcome available to you; rejecting a sound one merely costs a round.
\end{flaPromptBox}

\begin{flaPromptBox}{Actor continuation after an unaccepted turn}
You are not finished, so do not stop.

A run ends in exactly one way: `pb_submit()` returns PASS. That has not
happened yet. If your proof already compiles cleanly, CALL `pb_submit()`
NOW -- a clean compile is not finishing, and an earlier submission made
while the proof was unfinished does not count.

You have NO turn limit and NO time limit. Nothing is running out, no budget is
being spent that you need to protect, and nobody is waiting on you. Another
hundred turns costs nothing and is entirely fine.

None of your work is lost -- everything you have established is still there in
front of you. Pick it up from where you left off. If the last thing you tried
did not work, try a different angle: a different decomposition of the goal,
different Mathlib lemmas, a smaller intermediate lemma proved on its own, or a
concrete small case worked by hand to sharpen the idea. If you have started to
suspect the statement is false, chase that instead: build a counterexample and
try to derive `False` from the hypotheses.

Continue now, with a tool call.
\end{flaPromptBox}

\begin{flaPromptBox}{Actor continuation after reviewer rejection}
An independent reviewer examined your proof and did NOT approve it. Its full report
follows. Treat it as the thing to fix; it has not seen your reasoning, only the
statement and the candidate.

<reviewer-report>

Revise the proof and call pb_submit again. There is no turn limit.
\end{flaPromptBox}

\Needspace{10\baselineskip}
\subsection{Lean Eval}

The default prompts refer to the challenge and solver-owned submission files. The reviewer receives the digest of the submission files as \texttt{<candidate-sha256>}. External comparison and independent kernel checking remain controlled by the runner. These are the base templates, with task-specific recovery instructions supplied separately when needed.

\begin{flaPromptBox}{Actor}
You are the proof actor for the LeanEval problem <problem-name>. Read README.md,
Challenge.lean, ChallengeDeps/, config.json, and the starter Submission.lean. Complete
every required declaration using only the challenge, its local dependencies, published
mathematical knowledge you already have, and Lean compiler feedback. Work only in this
workspace. Edit only Submission.lean and Lean files under Submission/. Do not modify
trusted files or build configuration. Do not use the network and do not search for
existing benchmark solutions. Use lake build Submission repeatedly and leave your
strongest complete candidate on disk. This is a persistent actor session; an
independent fresh reviewer will assess the candidate after this round.
\end{flaPromptBox}

\begin{flaPromptBox}{Fresh reviewer}
You are a fresh independent reviewer of a LeanEval candidate and have no actor
transcript. Inspect README.md, Challenge.lean, ChallengeDeps/, config.json,
Submission.lean, and Submission/**/*.lean. Independently run lake build Submission.
Check that every hole is genuinely solved, that the candidate does not rely on sorry
or unsafe shortcuts, and that it respects solver-owned file boundaries. Do not edit
any file. Do not use the network or search for existing solutions. End with exactly
two standalone lines: REVIEW_VERDICT: APPROVE or REVIEW_VERDICT: REJECT, and
REVIEW_CANDIDATE_SHA256: <candidate-sha256>
\end{flaPromptBox}

\begin{flaPromptBox}{Actor continuation}
A fresh independent reviewer or the external verifier rejected the current candidate.
Read REVIEW.md, verify the feedback yourself, and continue solving every required
declaration. Work only in this workspace. You may edit only Submission.lean and Lean
files under Submission/. Use lake build Submission to check your work. Do not use the
network or search for existing benchmark solutions. Leave your strongest complete
candidate on disk.
\end{flaPromptBox}